\catcode`@=11 % This allows us to modify PLAIN macros.
%
%%%%%%%%%%%%%%%%%%%%%%%%%%%%%%%%%%%%%%%%%%%%%%%%%%%%%%%%%%%%%%%%%%%%%%%%
%
%   I begin with fonts.
%

\font\fourteenrm=cmr10 scaled\magstep2
\font\twelverm=cmr10 scaled\magstep1
\font\ninerm=cmr9	     \font\sixrm=cmr6

\font\fourteenbf=cmbx10 scaled\magstep2
\font\twelvebf=cmbx10 scaled\magstep1
\font\ninebf=cmbx9	      \font\sixbf=cmbx6
\font\seventeeni=cmmi10 scaled\magstep3	    \skewchar\seventeeni='177
\font\fourteeni=cmmi10 scaled\magstep2	    \skewchar\fourteeni='177
\font\twelvei=cmmi10 scaled\magstep1	    \skewchar\twelvei='177
\font\ninei=cmmi9			    \skewchar\ninei='177
\font\sixi=cmmi6			    \skewchar\sixi='177
\font\seventeensy=cmsy10 scaled\magstep3    \skewchar\seventeensy='60
\font\fourteensy=cmsy10 scaled\magstep2	    \skewchar\fourteensy='60
\font\twelvesy=cmsy10 scaled\magstep1	    \skewchar\twelvesy='60
\font\ninesy=cmsy9			    \skewchar\ninesy='60
\font\sixsy=cmsy6			    \skewchar\sixsy='60

\font\fourteenex=cmex10 scaled\magstep2
\font\twelveex=cmex10 scaled\magstep1

\font\fourteensl=cmsl10 scaled\magstep2
\font\twelvesl=cmsl10 scaled\magstep1
\font\ninesl=cmsl9

\font\fourteenit=cmti10 scaled\magstep2
\font\twelveit=cmti10 scaled\magstep1
\font\twelvett=cmtt10 scaled\magstep1
\font\twelvecp=cmcsc10 scaled\magstep1
\font\tencp=cmcsc10
\newfam\cpfam
%
	% quick fix for a missing font
%
\newcount\f@ntkey	     \f@ntkey=0
\def\samef@nt{\relax \ifcase\f@ntkey \rm \or\oldstyle \or\or
	 \or\it \or\sl \or\bf \or\tt \or\caps \fi }
\def\fourteenpoint{\relax
    \textfont0=\fourteenrm	    \scriptfont0=\tenrm
    \scriptscriptfont0=\sevenrm
     \def\rm{\fam0 \fourteenrm \f@ntkey=0 }\relax
    \textfont1=\fourteeni	    \scriptfont1=\teni
    \scriptscriptfont1=\seveni
     \def\oldstyle{\fam1 \fourteeni\f@ntkey=1 }\relax
    \textfont2=\fourteensy	    \scriptfont2=\tensy
    \scriptscriptfont2=\sevensy
    \textfont3=\fourteenex     \scriptfont3=\fourteenex
    \scriptscriptfont3=\fourteenex
    \def\it{\fam\itfam \fourteenit\f@ntkey=4 }\textfont\itfam=\fourteenit
    \def\sl{\fam\slfam \fourteensl\f@ntkey=5 }\textfont\slfam=\fourteensl
    \scriptfont\slfam=\tensl
    \def\bf{\fam\bffam \fourteenbf\f@ntkey=6 }\textfont\bffam=\fourteenbf
    \scriptfont\bffam=\tenbf	 \scriptscriptfont\bffam=\sevenbf
    \def\tt{\fam\ttfam \twelvett \f@ntkey=7 }\textfont\ttfam=\twelvett
    \h@big=11.9\p@{} \h@Big=16.1\p@{} \h@bigg=20.3\p@{} \h@Bigg=24.5\p@{}
    \def\caps{\fam\cpfam \twelvecp \f@ntkey=8 }\textfont\cpfam=\twelvecp
    \setbox\strutbox=\hbox{\vrule height 12pt depth 5pt width\z@}
    \samef@nt}
\def\twelvepoint{\relax
    \textfont0=\twelverm	  \scriptfont0=\ninerm
    \scriptscriptfont0=\sixrm
     \def\rm{\fam0 \twelverm \f@ntkey=0 }\relax
    \textfont1=\twelvei		  \scriptfont1=\ninei
    \scriptscriptfont1=\sixi
     \def\oldstyle{\fam1 \twelvei\f@ntkey=1 }\relax
    \textfont2=\twelvesy	  \scriptfont2=\ninesy
    \scriptscriptfont2=\sixsy
    \textfont3=\twelveex	  \scriptfont3=\twelveex
    \scriptscriptfont3=\twelveex
    \def\it{\fam\itfam \twelveit \f@ntkey=4 }\textfont\itfam=\twelveit
    \def\sl{\fam\slfam \twelvesl \f@ntkey=5 }\textfont\slfam=\twelvesl
    \scriptfont\slfam=\ninesl
    \def\bf{\fam\bffam \twelvebf \f@ntkey=6 }\textfont\bffam=\twelvebf
    \scriptfont\bffam=\ninebf	  \scriptscriptfont\bffam=\sixbf
    \def\tt{\fam\ttfam \twelvett \f@ntkey=7 }\textfont\ttfam=\twelvett
    \h@big=10.2\p@{}
    \h@Big=13.8\p@{}
    \h@bigg=17.4\p@{}
    \h@Bigg=21.0\p@{}
    \def\caps{\fam\cpfam \twelvecp \f@ntkey=8 }\textfont\cpfam=\twelvecp
    \setbox\strutbox=\hbox{\vrule height 10pt depth 4pt width\z@}
    \samef@nt}
\def\tenpoint{\relax
    \textfont0=\tenrm	       \scriptfont0=\sevenrm
    \scriptscriptfont0=\fiverm
    \def\rm{\fam0 \tenrm \f@ntkey=0 }\relax
    \textfont1=\teni	       \scriptfont1=\seveni
    \scriptscriptfont1=\fivei
    \def\oldstyle{\fam1 \teni \f@ntkey=1 }\relax
    \textfont2=\tensy	       \scriptfont2=\sevensy
    \scriptscriptfont2=\fivesy
    \textfont3=\tenex	       \scriptfont3=\tenex
    \scriptscriptfont3=\tenex
    \def\it{\fam\itfam \tenit \f@ntkey=4 }\textfont\itfam=\tenit
    \def\sl{\fam\slfam \tensl \f@ntkey=5 }\textfont\slfam=\tensl
    \def\bf{\fam\bffam \tenbf \f@ntkey=6 }\textfont\bffam=\tenbf
    \scriptfont\bffam=\sevenbf	   \scriptscriptfont\bffam=\fivebf
    \def\tt{\fam\ttfam \tentt \f@ntkey=7 }\textfont\ttfam=\tentt
    \def\caps{\fam\cpfam \tencp \f@ntkey=8 }\textfont\cpfam=\tencp
    \setbox\strutbox=\hbox{\vrule height 8.5pt depth 3.5pt width\z@}
    \samef@nt}
%
%%%%%%%%%%%%%%%%%%%%%%%%%%%%%%%%%%%%%%%%%%%%%%%%%%%%%%%%%%%%%%%%%%%%%%%%
%
%   Next redifine \big \Big \bigg and \Bigg to work with all fonts
%
%%%%%%%%%%%%%%%%%%%%%%%%%%%%%%%%%%%%%%%%%%%%%%%%%%%%%%%%%%%%%%%%%%%%%%%%
%
\newdimen\h@big  \h@big=8.5\p@
\newdimen\h@Big  \h@Big=11.5\p@
\newdimen\h@bigg  \h@bigg=14.5\p@
\newdimen\h@Bigg  \h@Bigg=17.5\p@
\def\big#1{{\hbox{$\left#1\vbox to\h@big{}\right.\n@space$}}}
\def\Big#1{{\hbox{$\left#1\vbox to\h@Big{}\right.\n@space$}}}
\def\bigg#1{{\hbox{$\left#1\vbox to\h@bigg{}\right.\n@space$}}}
\def\Bigg#1{{\hbox{$\left#1\vbox to\h@Bigg{}\right.\n@space$}}}
%
%%%%%%%%%%%%%%%%%%%%%%%%%%%%%%%%%%%%%%%%%%%%%%%%%%%%%%%%%%%%%%%%%%%%%%%%
%
%   Next, I define basic spacing parameters.
%
\normalbaselineskip = 20pt plus 0.2pt minus 0.1pt
\normallineskip = 1.5pt plus 0.1pt minus 0.1pt
\normallineskiplimit = 1.5pt
\newskip\normaldisplayskip
\normaldisplayskip = 20pt plus 5pt minus 10pt
\newskip\normaldispshortskip
\normaldispshortskip = 6pt plus 5pt
\newskip\normalparskip
\normalparskip = 6pt plus 2pt minus 1pt
\newskip\skipregister
\skipregister = 5pt plus 2pt minus 1.5pt
\newif\ifsingl@	   \newif\ifdoubl@
\newif\iftwelv@	   \twelv@true
\def\singlespace{\singl@true\doubl@false\spaces@t}
\def\doublespace{\singl@false\doubl@true\spaces@t}
\def\normalspace{\singl@false\doubl@false\spaces@t}
\def\Tenpoint{\tenpoint\twelv@false\spaces@t}
\def\Twelvepoint{\twelvepoint\twelv@true\spaces@t}
\def\spaces@t{\relax%
 \iftwelv@ \ifsingl@\subspaces@t3:4;\else\subspaces@t1:1;\fi%
 \else \ifsingl@\subspaces@t3:5;\else\subspaces@t4:5;\fi \fi%
 \ifdoubl@ \multiply\baselineskip by 5%
 \divide\baselineskip by 4 \fi \unskip}
\def\subspaces@t#1:#2;{%
      \baselineskip = \normalbaselineskip%
      \multiply\baselineskip by #1 \divide\baselineskip by #2%
      \lineskip = \normallineskip%
      \multiply\lineskip by #1 \divide\lineskip by #2%
      \lineskiplimit = \normallineskiplimit%
      \multiply\lineskiplimit by #1 \divide\lineskiplimit by #2%
      \parskip = \normalparskip%
      \multiply\parskip by #1 \divide\parskip by #2%
      \abovedisplayskip = \normaldisplayskip%
      \multiply\abovedisplayskip by #1 \divide\abovedisplayskip by #2%
      \belowdisplayskip = \abovedisplayskip%
      \abovedisplayshortskip = \normaldispshortskip%
      \multiply\abovedisplayshortskip by #1%
	\divide\abovedisplayshortskip by #2%
      \belowdisplayshortskip = \abovedisplayshortskip%
      \advance\belowdisplayshortskip by \belowdisplayskip%
      \divide\belowdisplayshortskip by 2%
      \smallskipamount = \skipregister%
      \multiply\smallskipamount by #1 \divide\smallskipamount by #2%
      \medskipamount = \smallskipamount \multiply\medskipamount by 2%
      \bigskipamount = \smallskipamount \multiply\bigskipamount by 4 }
\def\normalbaselines{ \baselineskip=\normalbaselineskip%
   \lineskip=\normallineskip \lineskiplimit=\normallineskip%
   \iftwelv@\else \multiply\baselineskip by 4 \divide\baselineskip by 5%
     \multiply\lineskiplimit by 4 \divide\lineskiplimit by 5%
     \multiply\lineskip by 4 \divide\lineskip by 5 \fi }
\Twelvepoint  % That's the default
\interlinepenalty=50
\interfootnotelinepenalty=5000
\predisplaypenalty=9000
\postdisplaypenalty=500
\hfuzz=1pt
\vfuzz=0.2pt
%
%%%%%%%%%%%%%%%%%%%%%%%%%%%%%%%%%%%%%%%%%%%%%%%%%%%%%%%%%%%%%%%%%%%%%%%%
%
%   Next, I define output routines, footnotes & related stuff.
%
\def\pagecontents{%
   \ifvoid\topins\else\unvbox\topins\vskip\skip\topins\fi
   \dimen@ = \dp255 \unvbox255
   \ifvoid\footins\else\vskip\skip\footins\footrule\unvbox\footins\fi
   \ifr@ggedbottom \kern-\dimen@ \vfil \fi }
\def\makeheadline{\vbox to 0pt{ \skip@=\topskip
      \advance\skip@ by -12pt \advance\skip@ by -2\normalbaselineskip
      \vskip\skip@ \line{\vbox to 12pt{}\the\headline} \vss
      }\nointerlineskip}
\def\makefootline{\baselineskip = 1.5\normalbaselineskip
		 \line{\the\footline}}
\newif\iffrontpage
\newif\ifletterstyle
\newif\ifp@genum
\def\nopagenumbers{\p@genumfalse}
\def\pagenumbers{\p@genumtrue}
\pagenumbers
\newtoks\paperheadline
\newtoks\letterheadline
\newtoks\letterfrontheadline
\newtoks\lettermainheadline
\newtoks\paperfootline
\newtoks\letterfootline
\newtoks\date
\footline={\ifletterstyle\the\letterfootline\else\the\paperfootline\fi}
\paperfootline={\hss\iffrontpage\else\ifp@genum\tenrm\folio\hss\fi\fi}
\letterfootline={\hfil}
\headline={\ifletterstyle\the\letterheadline\else\the\paperheadline\fi}
\paperheadline={\hfil}
\letterheadline{\iffrontpage\the\letterfrontheadline
     \else\the\lettermainheadline\fi}
\lettermainheadline={\rm\ifp@genum page \ \folio\fi\hfil\the\date}
\def\monthname{\relax\ifcase\month 0/\or January\or February\or
   March\or April\or May\or June\or July\or August\or September\or
   October\or November\or December\else\number\month/\fi}
\date={\monthname\ \number\day, \number\year}

\countdef\pagenumber=1  \pagenumber=1
\def\advancepageno{\global\advance\pageno by 1
   \ifnum\pagenumber<0 \global\advance\pagenumber by -1
    \else\global\advance\pagenumber by 1 \fi \global\frontpagefalse }
\def\folio{\ifnum\pagenumber<0 \romannumeral-\pagenumber
	   \else \number\pagenumber \fi }
\def\footrule{\dimen@=\prevdepth\nointerlineskip
   \vbox to 0pt{\vskip -0.25\baselineskip \hrule width 0.35\hsize \vss}
   \prevdepth=\dimen@ }
\newtoks\foottokens
\foottokens={\Tenpoint\singlespace}
\newdimen\footindent
\footindent=24pt
\def\vfootnote#1{\insert\footins\bgroup  \the\foottokens
   \interlinepenalty=\interfootnotelinepenalty \floatingpenalty=20000
   \splittopskip=\ht\strutbox \boxmaxdepth=\dp\strutbox
   \leftskip=\footindent \rightskip=\z@skip
   \parindent=0.5\footindent \parfillskip=0pt plus 1fil
   \spaceskip=\z@skip \xspaceskip=\z@skip
   \Textindent{$ #1 $}\footstrut\futurelet\next\fo@t}
\def\Textindent#1{\noindent\llap{#1\enspace}\ignorespaces}
\def\footnote#1{\attach{#1}\vfootnote{#1}}

\let\footsymbol=\star
\newcount\lastf@@t	     \lastf@@t=-1
\newcount\footsymbolcount    \footsymbolcount=0
\newif\ifPhysRev
%%%%%%%%%%%%%%%%%%%%%%%%%%%%%%%%%%%%%%%%%%%%%%%%%%%%%%%%%%%%%%%%%%%%%
% modified by S. Yoro. Reference for PTP June, 1987
\newif\ifP@TP
\def\Prog{\P@TPtrue}
\def\refdelim{\ifP@TP )\thinspace\fi}
\def\refoutitem{\ifP@TP\else .\fi}
%%%%%%%%%%%%%%%%%%%%%%%%%%%%%%%%%%%%%%%%%%%%%%%%%%%%%%%%%%%%%%%%%
%
\def\footsymbolgen{\relax \ifPhysRev \iffrontpage \NPsymbolgen\else
      \PRsymbolgen\fi \else \NPsymbolgen\fi
   \global\lastf@@t=\pageno \footsymbol }
\def\NPsymbolgen{\ifnum\footsymbolcount<0 \global\footsymbolcount=0\fi
   {\iffrontpage \else \advance\lastf@@t by 1 \fi
    \ifnum\lastf@@t<\pageno \global\footsymbolcount=0
     \else \global\advance\footsymbolcount by 1 \fi }
   \ifcase\footsymbolcount \fd@f\star\or \fd@f\dagger\or \fd@f\ast\or
    \fd@f\ddagger\or \fd@f\natural\or \fd@f\diamond\or \fd@f\bullet\or
    \fd@f\nabla\else \fd@f\dagger\global\footsymbolcount=0 \fi }
\def\fd@f#1{\xdef\footsymbol{#1}}
\def\PRsymbolgen{\ifnum\footsymbolcount>0 \global\footsymbolcount=0\fi
      \global\advance\footsymbolcount by -1
      \xdef\footsymbol{\sharp\number-\footsymbolcount} }
\def\space@ver#1{\let\@sf=\empty \ifmmode #1\else \ifhmode
   \edef\@sf{\spacefactor=\the\spacefactor}\unskip${}#1$\relax\fi\fi}
\def\attach#1{\space@ver{\strut^{\mkern 2mu #1} }\@sf\ }
%
%%%%%%%%%%%%%%%%%%%%%%%%%%%%%%%%%%%%%%%%%%%%%%%%%%%%%%%%%%%%%%%%%%%%%%%%
%
%   Here come chapter, section, subsection & appendix macros.
%
\newcount\chapternumber	     \chapternumber=0
\newcount\sectionnumber	     \sectionnumber=0
\newcount\equanumber	     \equanumber=0
\let\chapterlabel=0
\newtoks\chapterstyle	     \chapterstyle={\Number}
\newskip\chapterskip	     \chapterskip=\bigskipamount
\newskip\sectionskip	     \sectionskip=\medskipamount
\newskip\headskip	     \headskip=8pt plus 3pt minus 3pt
\newdimen\chapterminspace    \chapterminspace=15pc
\newdimen\sectionminspace    \sectionminspace=10pc
\newdimen\referenceminspace  \referenceminspace=25pc
\def\chapterreset{\global\advance\chapternumber by 1
   \ifnum\equanumber<0 \else\global\equanumber=0\fi
   \sectionnumber=0 \makel@bel}
\def\makel@bel{\xdef\chapterlabel{%
\the\chapterstyle{\the\chapternumber}.}}
\def\sectionlabel{\number\sectionnumber \quad }
\def\alphabetic#1{\count255='140 \advance\count255 by #1\char\count255}
\def\Alphabetic#1{\count255='100 \advance\count255 by #1\char\count255}
\def\Roman#1{\uppercase\expandafter{\romannumeral #1}}
\def\roman#1{\romannumeral #1}
\def\Number#1{\number #1}
\def\unnumberedchapters{\let\makel@bel=\relax \let\chapterlabel=\relax
\let\sectionlabel=\relax \equanumber=-1 }
\def\titlestyle#1{\par\begingroup \interlinepenalty=9999
     \leftskip=0.02\hsize plus 0.23\hsize minus 0.02\hsize
     \rightskip=\leftskip \parfillskip=0pt
     \hyphenpenalty=9000 \exhyphenpenalty=9000
     \tolerance=9999 \pretolerance=9000
     \spaceskip=0.333em \xspaceskip=0.5em
     \iftwelv@\fourteenpoint\else\twelvepoint\fi
   \noindent #1\par\endgroup }
\def\spacecheck#1{\dimen@=\pagegoal\advance\dimen@ by -\pagetotal
   \ifdim\dimen@<#1 \ifdim\dimen@>0pt \vfil\break \fi\fi}
\def\chapter#1{\par \penalty-300 \vskip\chapterskip
   \spacecheck\chapterminspace
   \chapterreset \titlestyle{\chapterlabel \ #1}
   \nobreak\vskip\headskip \penalty 30000
   \wlog{\string\chapter\ \chapterlabel} }

\def\section#1{\par \ifnum\the\lastpenalty=30000\else
   \penalty-200\vskip\sectionskip \spacecheck\sectionminspace\fi
   \wlog{\string\section\ \chapterlabel \the\sectionnumber}
   \global\advance\sectionnumber by 1  \noindent
   {\caps\enspace\chapterlabel \sectionlabel #1}\par
   \nobreak\vskip\headskip \penalty 30000 }
\def\subsection#1{\par
   \ifnum\the\lastpenalty=30000\else \penalty-100\smallskip \fi
   \noindent\undertext{#1}\enspace \vadjust{\penalty5000}}

\def\undertext#1{\vtop{\hbox{#1}\kern 1pt \hrule}}
\def\ack{\par\penalty-100\medskip \spacecheck\sectionminspace
   \line{\fourteenrm\hfil ACKNOWLEDGEMENTS\hfil}\nobreak\vskip\headskip }
\def\APPENDIX#1#2{\par\penalty-300\vskip\chapterskip
   \spacecheck\chapterminspace \chapterreset \xdef\chapterlabel{#1}
   \titlestyle{APPENDIX #2} \nobreak\vskip\headskip \penalty 30000
   \wlog{\string\Appendix\ \chapterlabel} }
\def\Appendix#1{\APPENDIX{#1}{#1}}
\def\appendix{\APPENDIX{A}{}}
%
%%%%%%%%%%%%%%%%%%%%%%%%%%%%%%%%%%%%%%%%%%%%%%%%%%%%%%%%%%%%%%%%%%%%%%%%
%
%   Here come macros for equation numbering.
%
\def\eqname#1{\relax \ifnum\equanumber<0
     \xdef#1{{\rm(\number-\equanumber)}}\global\advance\equanumber by -1
    \else \global\advance\equanumber by 1
      \xdef#1{{\rm(\chapterlabel \number\equanumber)}} \fi}
\def\eqinsert#1{\noalign{\dimen@=\prevdepth \nointerlineskip
   \setbox0=\hbox to\displaywidth{\hfil #1}
   \vbox to 0pt{\vss\hbox{$\!\box0\!$}\kern-0.5\baselineskip}
   \prevdepth=\dimen@}}
%

%

%

%
%%%%%%%%%%%%%%%%%%%%%%%%%%%%%%%%%%%%%%%%%%%%%%%%%%%%%%%%%%%%%%%%%%%%%%%%
%   Here come items and lists
%
\def\GENITEM#1;#2{\par \hangafter=0 \hangindent=#1
    \Textindent{$ #2 $}\ignorespaces}
\outer\def\newitem#1=#2;{\gdef#1{\GENITEM #2;}}
\newdimen\itemsize		  \itemsize=30pt
\newitem\item=1\itemsize;
\newitem\sitem=1.75\itemsize;	  
\newitem\ssitem=2.5\itemsize;	  
\outer\def\newlist#1=#2&#3&#4;{\toks0={#2}\toks1={#3}%
   \count255=\escapechar \escapechar=-1
   \alloc@0\list\countdef\insc@unt\listcount	 \listcount=0
   \edef#1{\par
      \countdef\listcount=\the\allocationnumber
      \advance\listcount by 1
      \hangafter=0 \hangindent=#4
      \Textindent{\the\toks0{\listcount}\the\toks1}}
   \expandafter\expandafter\expandafter
    \edef\c@t#1{begin}{\par
      \countdef\listcount=\the\allocationnumber \listcount=1
      \hangafter=0 \hangindent=#4
      \Textindent{\the\toks0{\listcount}\the\toks1}}
   \expandafter\expandafter\expandafter
    \edef\c@t#1{con}{\par \hangafter=0 \hangindent=#4 \noindent}
   \escapechar=\count255}
\def\c@t#1#2{\csname\string#1#2\endcsname}
\newlist\point=\Number&.&1.0\itemsize;
\newlist\subpoint=(\alphabetic&)&1.75\itemsize;
\newlist\subsubpoint=(\roman&)&2.5\itemsize;
%

%
%%%%%%%%%%%%%%%%%%%%%%%%%%%%%%%%%%%%%%%%%%%%%%%%%%%%%%%%%%%%%%%%%%%%%%%%
%
%   Here come macros for references, figures & tables.
%
\newcount\referencecount     \referencecount=0
\newif\ifreferenceopen	     \newwrite\referencewrite
\newtoks\rw@toks
%%%%%%%%%%%%%%%%%%%%%%%%%%%%%%%%%%%%%%%%%%%%%%%%%%%%%%%%%%%%%%
% modified by S. Yoro.   June, 1987
\def\NPrefmark#1{\attach{[ #1 ]}}
\let\PTPrefmark=\attach
%%%%%%%%%%%%%%%%%%%%%%%%%%%%%%%%%%%%%%%%%%%%%%%%%%%%%%%%%%%%%%
%
\let\PRrefmark=\attach
%
%%%%%%%%%%%%%%%%%%%%%%%%%%%%%%%%%%%%%%%%%%%%%%%%%%%%%%%%%%%%%
%   modified by S. Yoro. June 1987
\def\refmark#1{\relax\ifPhysRev\PRrefmark{#1}\else
       {\relax\ifP@TP\PTPrefmark{#1}\else\NPrefmark{#1}\fi}\fi}
%%%%%%%%%%%%%%%%%%%%%%%%%%%%%%%%%%%%%%%%%%%%%%%%%%%%%%%%%%%%%%
%
\def\refend{\refmark{\number\referencecount\refdelim}}
\newcount\lastrefsbegincount \lastrefsbegincount=0
\def\refsend{\refmark{\count255=\referencecount
   \advance\count255 by-\lastrefsbegincount
%%%%%%%%%%%%%%%%%%%%%%%%%%%%%%%%%%%%%%%%%%%%%%%%
%   modified by S. Yoro.   June 1987
    \ifcase\count255 \number\referencecount\refdelim
   \or \number\lastrefsbegincount\refdelim ,\number\referencecount\refdelim
   \else \number\lastrefsbegincount\refdelim -\number\referencecount\refdelim
\fi}}
%%%%%%%%%%%%%%%%%%%%%%%%%%%%%%%%%%%%%%%%%%%%%%%%%%%%%%%%
%
\def\refch@ck{\chardef\rw@write=\referencewrite
   \ifreferenceopen \else \referenceopentrue
   \immediate\openout\referencewrite=reference.aux \fi}
%
% In \obeyendofline, we say `\let^^M=\relax
{\catcode`\^^M=\active % these lines must end with %
  \gdef\obeyendofline{\catcode`\^^M\active \let^^M\ }}%
%
% In \ignoreendofline, we say `\let^^M=\relax
{\catcode`\^^M=\active % these lines must end with %
  \gdef\ignoreendofline{\catcode`\^^M=5}}
{\obeyendofline\gdef\rw@start#1{\def\t@st{#1} \ifx\t@st\blankend%
\endgroup \@sf \relax \else \ifx\t@st\bl@nkend \endgroup \@sf \relax%
\else \rw@begin#1
\backtotext
\fi \fi } }
{\obeyendofline\gdef\rw@begin#1
{\def\n@xt{#1}\rw@toks={#1}\relax%
\rw@next}}
\def\blankend{}
{\obeylines\gdef\bl@nkend{
}}
\newif\iffirstrefline  \firstreflinetrue
\def\rwr@teswitch{\ifx\n@xt\blankend \let\n@xt=\rw@begin %
 \else\iffirstrefline \global\firstreflinefalse%
\immediate\write\rw@write{\noexpand\obeyendofline \the\rw@toks}%
\let\n@xt=\rw@begin%
      \else\ifx\n@xt\rw@@d \def\n@xt{\immediate\write\rw@write{%
	\noexpand\ignoreendofline}\endgroup \@sf}%
	     \else \immediate\write\rw@write{\the\rw@toks}%
	     \let\n@xt=\rw@begin\fi\fi \fi}
\def\rw@next{\rwr@teswitch\n@xt}
\def\rw@@d{\backtotext} \let\rw@end=\relax
\let\backtotext=\relax

\newdimen\refindent	\refindent=30pt
\def\refitem#1{\par \hangafter=0 \hangindent=\refindent \Textindent{#1}}
%%%%%%%%%%%%%%%%%%%%%%%%%%%%%%%%%%%%%%%%%%%%%%%%%%%%%%%
%Modified by S. Yoro.    June ,1987
%%%%%%%%%%%%%%%%%%%%%%%%%%%%%%%%%%%%%%%%%%%%%%%%%%%%%%%
\def\REFNUM#1{\space@ver{}\refch@ck \firstreflinetrue%
 \global\advance\referencecount by 1 \xdef#1{\the\referencecount\refdelim}}
\def\refnum#1{\space@ver{}\refch@ck \firstreflinetrue%
 \global\advance\referencecount by 1
\xdef#1{\the\referencecount\refdelim}\refend}

\def\REF#1{\REFNUM#1%
 \immediate\write\referencewrite{%
 \noexpand\refitem{#1\refoutitem}}%
\begingroup\obeyendofline\rw@start}
\def\ref{\refnum\?%
 \immediate\write\referencewrite{\noexpand\refitem{\?\refoutitem}}%
\begingroup\obeyendofline\rw@start}
\def\Ref#1{\refnum#1%
 \immediate\write\referencewrite{\noexpand\refitem{#1\refoutitem}}%
\begingroup\obeyendofline\rw@start}
\def\REFS#1{\REFNUM#1\global\lastrefsbegincount=\referencecount
\immediate\write\referencewrite{\noexpand\refitem{#1\refoutitem}}%
\begingroup\obeyendofline\rw@start}
%%%%%%%%%%%%%%%%%%%%%%%%%%%%%%%%%%%%%%%%%%%%%%%%%%%%%%%%%%%%%%%%%%%%%
%

%

%
\newcount\figurecount	  \figurecount=0
\newif\iffigureopen	  \newwrite\figurewrite
\def\figch@ck{\chardef\rw@write=\figurewrite \iffigureopen\else
   \immediate\openout\figurewrite=figures.aux
   \figureopentrue\fi}
\def\FIGNUM#1{\space@ver{}\figch@ck \firstreflinetrue%
 \global\advance\figurecount by 1 \xdef#1{\the\figurecount}}
\def\FIG#1{\FIGNUM#1
   \immediate\write\figurewrite{\noexpand\refitem{#1.}}%
   \begingroup\obeyendofline\rw@start}
\def\fig{\FIGNUM\? fig.~\?%
\immediate\write\figurewrite{\noexpand\refitem{\?.}}%
\begingroup\obeyendofline\rw@start}
\def\figure{\FIGNUM\? figure~\?
   \immediate\write\figurewrite{\noexpand\refitem{\?.}}%
   \begingroup\obeyendofline\rw@start}
\def\Fig{\FIGNUM\? Fig.~\?%
\immediate\write\figurewrite{\noexpand\refitem{\?.}}%
\begingroup\obeyendofline\rw@start}
\def\Figure{\FIGNUM\? Figure~\?%
\immediate\write\figurewrite{\noexpand\refitem{\?.}}%
\begingroup\obeyendofline\rw@start}
\newcount\tablecount	 \tablecount=0
\newif\iftableopen	 \newwrite\tablewrite
\def\tabch@ck{\chardef\rw@write=\tablewrite \iftableopen\else
   \immediate\openout\tablewrite=tables.aux
   \tableopentrue\fi}
\def\TABNUM#1{\space@ver{}\tabch@ck \firstreflinetrue%
 \global\advance\tablecount by 1 \xdef#1{\the\tablecount}}
\def\TABLE#1{\TABNUM#1
   \immediate\write\tablewrite{\noexpand\refitem{#1.}}%
   \begingroup\obeyendofline\rw@start}
\def\Table{\TABNUM\? Table~\?%
\immediate\write\tablewrite{\noexpand\refitem{\?.}}%
\begingroup\obeyendofline\rw@start}
%

%
%%%%%%%%%%%%%%%%%%%%%%%%%%%%%%%%%%%%%%%%%%%%%%%%%%%%%%%%%%%%%%%%%%%%%%%%
%
%   Here come macros for memos & letters.
%
\def\masterreset{\global\pagenumber=1 \global\chapternumber=0
   \global\equanumber=0 \global\sectionnumber=0
   \global\referencecount=0 \global\figurecount=0 \global\tablecount=0 }
\def\FRONTPAGE{\ifvoid255\else\vfill\penalty-2000\fi
      \masterreset\global\frontpagetrue
      \global\lastf@@t=0 \global\footsymbolcount=0}

\def\paperstyle{\letterstylefalse\normalspace\papersize}
\def\letterstyle{\letterstyletrue\singlespace\lettersize}
%
%\def\papersize{\hsize=35pc\vsize=50pc\hoffset=1pc\voffset=6pc
%		\skip\footins=\bigskipamount}
% 6/24/1986 two lines above was modified by J.CHIBA as below
\def\papersize{\hsize=35pc\vsize=50pc\hoffset=1cm\voffset=0pc
	       \skip\footins=\bigskipamount}
%\def\lettersize{\hsize=6.5in\vsize=8.5in\hoffset=0in\voffset=1in
% 6/25/1986 The above was modified as below by H.Mawatari
%\def\lettersize{\hsize=6.5in\vsize=8.5in\hoffset=0.48in\voffset=1in
%   \skip\footins=\smallskipamount \multiply\skip\footins by 3 }
%%%%%%%%%%%%%%%%%%%%%%%%%%%%%%%%%%%%%%%%%%%%%%%%%%%%%%%%%
% modified by K. Suehiro July 1988
\def\lettersize{\hsize=35pc\vsize=50pc\hoffset=1cm\voffset=0pc
   \skip\footins=\smallskipamount \multiply\skip\footins by 3 }
\paperstyle   %  This is the default
%
% % % % % % % % % % % % % % % % % % % % % % % % % % % % % % % % % % % %
%
\def\MEMO{\letterstyle\FRONTPAGE \letterfrontheadline={\hfil}
    \line{\quad\fourteenrm MEMORANDUM\hfil\twelverm\the\date\quad}
    \medskip \memod@f}

\def\memit@m#1{\smallskip \hangafter=0 \hangindent=1in
      \Textindent{\caps #1}}
\def\memod@f{\xdef\to{\memit@m{To:}}\xdef\from{\memit@m{From:}}%
     \xdef\topic{\memit@m{Topic:}}\xdef\subject{\memit@m{Subject:}}%
     \xdef\rule{\bigskip\hrule height 1pt\bigskip}}
\memod@f
\newskip\lettertopfil
\lettertopfil = 0pt plus 1.5in minus 0pt
\newskip\letterbottomfil
\letterbottomfil = 0pt plus 2.3in minus 0pt
\newskip\spskip \setbox0\hbox{\ } \spskip=-1\wd0
\def\addressee#1{\medskip\rightline{\the\date\hskip 30pt} \bigskip
   \vskip\lettertopfil
   \ialign to\hsize{\strut ##\hfil\tabskip 0pt plus \hsize \cr #1\crcr}
   \medskip\noindent\hskip\spskip}
\newskip\signatureskip	     \signatureskip=40pt
\def\signed#1{\par \penalty 9000 \bigskip \dt@pfalse
  \everycr={\noalign{\ifdt@p\vskip\signatureskip\global\dt@pfalse\fi}}
  \setbox0=\vbox{\singlespace \halign{\tabskip 0pt \strut ##\hfil\cr
   \noalign{\global\dt@ptrue}#1\crcr}}
  \line{\hskip 0.5\hsize minus 0.5\hsize \box0\hfil} \medskip }

\def\endletter{\ifnum\pagenumber=1 \vskip\letterbottomfil\supereject
\else \vfil\supereject \fi}
\newbox\letterb@x
\def\lettertext{\par\unvcopy\letterb@x\par}
\def\multiletter{\setbox\letterb@x=\vbox\bgroup
      \everypar{\vrule height 1\baselineskip depth 0pt width 0pt }
      \singlespace \topskip=\baselineskip }
\def\letterend{\par\egroup}
%
%%%%%%%%%%%%%%%%%%%%%%%%%%%%%%%%%%%%%%%%%%%%%%%%%%%%%%%%%%%%%%%%%%%%%%%
%
%   Here come macros for title pages.
%
\newskip\frontpageskip
\newtoks\pubtype
\newtoks\Pubnum
\newtoks\pubnum
\newif\ifp@bblock  \p@bblocktrue
\def\PH@SR@V{\doubl@true \baselineskip=24.1pt plus 0.2pt minus 0.1pt
	     \parskip= 3pt plus 2pt minus 1pt }
\def\PHYSREV{\paperstyle\PhysRevtrue\PH@SR@V}
\def\titlepage{\FRONTPAGE\paperstyle\ifPhysRev\PH@SR@V\fi
   \ifp@bblock\p@bblock\fi}
\def\nopubblock{\p@bblockfalse}
\def\endpage{\vfil\break}
\frontpageskip=1\medskipamount plus .5fil
\pubtype={ }
\newtoks\publevel
\publevel={Report}   % The alternatives are Internal and Preprint
\Pubnum={}
%\pubnum={0000}
%
\def\p@bblock{\begingroup \tabskip=\hsize minus \hsize
   \baselineskip=1.5\ht\strutbox \topspace-2\baselineskip
   \halign to\hsize{\strut ##\hfil\tabskip=0pt\crcr
   \the\Pubnum\cr \the\date\cr }\endgroup}
\def\title#1{\vskip\frontpageskip \titlestyle{#1} \vskip\headskip }
\def\author#1{\vskip\frontpageskip\titlestyle{\twelvecp #1}\nobreak}

\def\address#1{\par\kern 5pt\titlestyle{\twelvepoint\it #1}}
\def\andaddress{\par\kern 5pt \centerline{\sl and} \address}

% 6/24/1986 two lines below were added by H.Mawatari

%
\def\abstract{\vskip\frontpageskip\centerline{\fourteenrm ABSTRACT}
	      \vskip\headskip }

%
%
%%%%%%%%%%%%%%%%%%%%%%%%%%%%%%%%%%%%%%%%%%%%%%%%%%%%%%%%%%%%%%%%%%%%%%%%
%   Miscellaneous macros
%

\def\\{\relax\ifmmode\backslash\else$\backslash$\fi}
\def\globaleqnumbers{\relax\if\equanumber<0\else\global\equanumber=-1\fi}
\def\nextline{\unskip\nobreak\hskip\parfillskip\break}

\def\journal#1&#2(#3){\unskip, \sl #1~\bf #2 \rm (19#3) }

\def\topspace{\hrule height 0pt depth 0pt \vskip}

\def\VEV#1{\left\langle #1\right\rangle}

\def\prop{\mathrel{{\mathchoice{\pr@p\scriptstyle}{\pr@p\scriptstyle}{
		\pr@p\scriptscriptstyle}{\pr@p\scriptscriptstyle} }}}
\def\pr@p#1{\setbox0=\hbox{$\cal #1 \char'103$}
   \hbox{$\cal #1 \char'117$\kern-.4\wd0\box0}}
\def\lsim{\mathrel{\mathpalette\@versim<}}
\def\gsim{\mathrel{\mathpalette\@versim>}}
\def\@versim#1#2{\lower0.2ex\vbox{\baselineskip\z@skip\lineskip\z@skip
  \lineskiplimit\z@\ialign{$\m@th#1\hfil##\hfil$\crcr#2\crcr\sim\crcr}}}
%
% % % % % % % % % % % % % % % % % % % % % % % % % % % % % % % % % % % %
%
%   Finally, some bug fixings.
%
\let\sec@nt=\sec
\def\sec{\relax\ifmmode\let\n@xt=\sec@nt\else\let\n@xt\section\fi\n@xt}
\def\obsolete#1{\message{Macro \string #1 is obsolete.}}
\def\firstsec#1{\obsolete\firstsec \section{#1}}
\def\firstsubsec#1{\obsolete\firstsubsec \subsection{#1}}
\def\thispage#1{\obsolete\thispage \global\pagenumber=#1\frontpagefalse}
\def\thischapter#1{\obsolete\thischapter \global\chapternumber=#1}
\def\nextequation#1{\obsolete\nextequation \global\equanumber=#1
   \ifnum\the\equanumber>0 \global\advance\equanumber by 1 \fi}
\def\BOXITEM{\afterassigment\B@XITEM\setbox0=}
\def\B@XITEM{\par\hangindent\wd0 \noindent\box0 }
%

%%%%%%%%%%%%%%%%%%%%%%%%%%%%%%%%%%%%%%%%%%%%%%%%%%%%%%%%%%%%%%%%%%%%%%%%
%   That's about it
%
\catcode`@=12 % at signs are no longer letters
\message{ by V.K.}
%

%%%%%%%%%%%%%%%%%%%%%%%%%%%%%%%%%%%%%%%%%%%%%%%%%%%%%%%%%%%%%%%%%%%%%%%%%%%%
%%%%%%
%%%%%% Up to here PHYZZX file
%%%%%%
%%%%%%%%%%%%%%%%%%%%%%%%%%%%%%%%%%%%%%%%%%%%%%%%%%%%%%%%%%%%%%%%%%%%%%%%%%%%

%\input particle
%
\def\sp(#1){ \noalign{\vskip #1pt} }
\font\BIGr=cmr10 scaled \magstep2
\Pubnum={
INS-Rep. 972
}
\date={Apr. 1993}
\titlepage
\title{\bf Fractal Structures of Quantum Gravity in Two Dimensions }
\author{ Noboru KAWAMOTO}
%Y. SAEKI$^{\dagger}$  and Y. WATABIKI }
%
\vskip 16pt
\address{     Institute for Nuclear Study, University of Tokyo,      \break
              Tanashi, Tokyo 188, Japan                              \break
}
%         $\dagger$ Hitachi,         \break
%                  , Japan }
%
%======================================================================%
%
%
%
%%%%%%%%%%%%%%%%%%%%%%%%%%%%%% ABSTRACT %%%%%%%%%%%%%%%%%%%%%%%%%%%%
%
\abstract{
Recent numerical results on the fractal structure of two-dimensional quantum
gravity coupled to $c=-2$ matter are reviewed.
Analytic derivation of the fractal dimensions based on the Liouville theory
and diffusion equation is also discussed.
Excellent agreements between the numerical and theoretical results are
obtained.
Some problems on the non-universal nature of the fractal structure in the
continuum limit are pointed out.
}
%
%%%%%%%%%%%%%%%%%%%%%%%%%%%%%%%%%%%%%%%%%%%%%%%%%%%%%%%%%%%%%%%%%%%%%%%%
%
\vskip 3cm
\noindent
-------------------------------------------------------------------------
---------------------
\nextline
Talk given at 7th Nishinomiya-Yukawa Memorial Symposium, held at \nextline
Nishinomiya-city on 18-20, Nov. 1992, to be published in the Proceedings.

\endpage             %The titlepage ends at this place.
%
%======================================================================%
%
%           REFERENCES
%
%\def\Comment#1{}

%
\REF\KPZDDK{
             V. G. Knizhnik, A. M. Polyakov and A. A. Zamolodchickov,
                Mod.\ Phys.\ Lett.\ {\bf A3} (1988) 819;
\nextline
             F. David,
                Mod.\ Phys.\ Lett.\ {\bf A3} (1988) 651;
\nextline
             J. Distler and H. Kawai,
                Nucl.\ Phys.\ {\bf B321} (1989) 509.
}
\REF\DT{
             F. David,
                Nucl.\ Phys.\ {\bf B257} (1985) 45;
\nextline
             V. A. Kazakov,
                Phys.\ Lett.\ {\bf 150B} (1985) 282;
\nextline
             D. V. Boulatov, V. A. Kazakov, I. K. Kostov and A. A. Migdal,
                Nucl.\ Phys.\ {\bf B275} (1986) 641;
\nextline
             J. Ambj\o rn, B. Durhuus and J. Fr\"ohlich,
                Nucl.\ Phys.\ {\bf B257} (1985) 433.
	     }
\REF\Matrix{
             E. Br\'ezin and V. A. Kazakov,
                Phys.\ Lett.\ {\bf 236B} (1990) 144;
\nextline
             M. Douglas and S. Shenker,
                Nucl.\ Phys.\ {\bf B335} (1990) 635;
\nextline
             D. J. Gross and A. A. Migdal,
                Phys.\ Rev.\ Lett.\ {\bf 64} (1990) 127.
            }
%\REF\BIPZ{
%             E. Br\'ezin, C. Itzykson, G. Parisi and J. B. Zuber,
%                Commun.\ Math.\ Phys.\ {\bf 59} (1978) 35.
%            }
\REF\AM{
             M. E. Agishtein and A. A. Migdal,
                Int.\ J.\ Mod.\ Phys.\ {\bf C1} (1990) 165;
                Nucl.\ Phys.\ {\bf B350} (1991) 690.
             }
\REF\KKSW{
             N. Kawamoto, V. A. Kazakov, Y. Saeki and Y. Watabiki,
                Phys.\ Rev.\ Lett.\ {\bf 68} (1992) 2113;
                Nucl.\ Phys.\ {\bf B}(Proc. Suppl.){\bf 26} (1992) 584.
             }
\REF\KN{
             H. Kawai and M. Ninomiya,
                Nucl.\ Phys.\ {\bf B336} (1990) 115.
	     }
\REF\D{
             F. David,
                Nucl.\ Phys.\ {\bf B368} (1992) 671.
	     }
\REF\KSW{
             N. Kawamoto, Y. Saeki and Y. Watabiki,
             in preparation.
	     }
\REF\ThreeDT{
             M. E. Agishtein and A. A. Migdal,
                Mod.\ Phys.\ Lett.\ {\bf A6} (1991) 1863;
\nextline
             J. Ambj\o rn and S. Varsted,
                Phys.\ Lett.\ {\bf 266B} (1991) 285;
\nextline
             D. V. Boulatov and A. Krzywicki,
                Mod.\ Phys.\ Lett.\ {\bf A6} (1991) 3005;
\nextline
             J. Ambj\o rn, D. V. Boulatov, A. Krzywicki and S. Varsted,
                Phys.\ Lett.\ {\bf 276B} (1992) 432.
	     }
\REF\FourDT{
             M. E. Agishtein and A. A. Migdal,
                Mod.\ Phys.\ Lett.\ {\bf A7} (1992) 1039;
\nextline
             J. Ambj\o rn and J. Jurkiewicz,
                Phys.\ Lett.\ {\bf 278B} (1992) 42.
	     }
\REF\KKMW{
             H. Kawai, N. Kawamoto, T. Mogami and Y. Watabiki,
             INS-Rep.-969, UT-633, Feb.1993, to be pub. in Phys. Lett.B.
             }
\REF\Kostov{
             V. A. Kazakov, I. K. Kostov and A. A. Migdal,
                Phys.\ Lett.\ {\bf 157B} 295 (1985);
\nextline
             D. V. Boulatov, V. A. Kazakov, I. K. Kostov and A. A. Migdal,
                Nucl.\ Phys.\ {\bf B275} 641 (1986);
\nextline
             I. K. Kostov and M. L. Mehta,
                Phys.\ Lett.\ {\bf 189B} 118 (1987).
	     }
\REF\TY{
             N. Tsuda and T. Yukawa, KEK preprint.
             }
\REF\KKMSW{
             H. Kawai, N. Kawamoto, T. Mogami, Y. Saeki and Y. Watabiki,
                in preparation.
             }
\def\Refmark(#1){$\,$[#1]}
\def\CMPrefmark(#1){ [#1]}

%
%======================================================================%
%
\topskip 30pt
\noindent
\fourteenrm
1.~Intoduction
\rm
\par
One of the most important recent developments in gravity theory is that we
obtained a well defined regularization of quantum gravity in two dimensions.
This recognition comes from the fact that the continuum
formulation\NPrefmark{\KPZDDK} and the dynamical triangulation\NPrefmark{\DT}
are equivalent.
The dynamical triangulation can be identified as a lattice regularization
of quantum gravity and can be analyzed analytically by the matrix
model\NPrefmark{\Matrix} and numerically by the computer
simulation.\NPrefmark{\AM,\KKSW}
It has become clear numerically that the fractal structure of space time
is the fundamental nature of the quantum gravity.\NPrefmark{\AM,\KKSW}
Analytical investigations by Liouville quantum gravity support the numerical
results.\NPrefmark{\KN,\D,\KSW}
It is an interesting question if the dynamical triangulation works as a
regularization scheme of three- and four-dimensional quantum gravity,
which may be tested only by numerical simulations at this
moment.\NPrefmark{\ThreeDT,\FourDT}
\par
In this manuscript we will summarize the recent numerical simulations
of two-dimensional quantum gravity coupled to $c=-2$
matter.\NPrefmark{\KKSW,\KSW}
We also show that our analytical investigations by Liouville
theory\NPrefmark{\KSW} have excellent agreements with the numerical
estimation of a fractal dimension and the mean squared distance of
gravitational random walks treated by a diffusion equation.
\par
Our recent analytic investigation on the fractal structure of $c=0$
model\NPrefmark{\KKMW} suggests that we must be careful about non-universal
behaviors of the fractal structures in the continuum limit of dynamical
triangulation.
\vskip 1cm

\noindent
\fourteenrm
2.~The shape of Typical Quantum Gravitational Space-Time and the Fractal
Structure
\rm
\par
In the dynamical triangulation of two-dimensional gravity, the metric
integration in the sense of path integral is replaced by a summation of
all the different types of surface configuration with a given number of
triangles and a given topology.
Here we specify the space-time topology as a sphere.
Formally the continuum partition function is given by
%%%%%%%%%%%%%%%%%%%%%%%%%%%%%%% Equation %%%%%%%%%%%%%%%%%%%%%%%%%%%%%
$$ \eqalign{ \sp (2.0)
 Z(A) \ = \ \int {\cal D}g ~\ \delta(\int dx\sqrt g -A) ~\, Z_m[g],
\cr
\sp(3.0)} \eqno(1) $$
%---------------------------------------------------------------------
where $Z_m[g]$ is a matter part of the partition function with gravitational
background and $A$ is the total area.
Reguralized counterpart of the above partition function by dynamical
triangulation is
%%%%%%%%%%%%%%%%%%%%%%%%%%%%%%% Equation %%%%%%%%%%%%%%%%%%%%%%%%%%%%%
$$ \eqalign{ \sp (2.0)
 Z_{reg}(A) \ = \ \sum_{G} \ Z_m[G] \, \delta_{Na^2,A}~~
              \sim \  Z_m[G_0],
\cr
\sp(3.0)} \eqno(2) $$
%---------------------------------------------------------------------
where $N$ is the number of equilateral triangles and $a^2$ is the area of
the triangle.
$G$ denotes a triangulation and $G_0$ is the typical triangulation which we
select from the huge set of triangulations.
The last approximate equality in Eq.(2) is valid up to the normalization
factor and if the selection of the typical surface is carried out by a
correct procedure which we explain later.
Since the path integration of the metric is carried out after the selection
of the typical surface, $G_0$ carries the information of the quantum
fluctuation of space time effectively.
\par
A natural question now is: \lq\lq How does the typical surface look like ?''
Since the typical surface carries the information of quantum gravitational
fluctuation, it may look quite different from the classical space time.
Some of the possible surfaces we can imagine beforehand may be; 1) smooth
surface, 2) spiky surface, 3) branched surface, ....
The next question is: \lq\lq How do we parametrize the shape of the typical
surface ?''
\par
Suppose we obtain the typical surface by the recursive sampling method which
we explain later, the actual measurements for the typical surface are carried
out as follows.
We first fix a marking site on the triangulated surface, from which we
measure the internal geodesic distance $r$.
We count the number of triangles ($V_t(r)$), links ($V_l(r)$), and sites
($V_s(r)$) within $r$ geodesic steps.
$V_t(r)$ is essentially the two-dimensional volume in radius $r$.
The number of the disconnected boundaries $N_b(r)$ can be counted by using
the following relation:
$\chi = 2 - N_b(r)$ and $\chi = V_s(r) - V_l(r) + V_t(r)$,
where $\chi$ is the Euler number of sphere with $N_b(r)$ boundary holes.
$L_b(r) = V_t(r+1) - V_t(r)$ is essentially a derivative of $V_t(r)$ with
respect to the geodesic distance and is the total length of the $N_b(r)$
disconnected boundaries at the distance $r$ from the marking site.
\par
In the actual calculations we define the geodesic distance on the
triangulated lattice.
To be precise, the region of geodesic distance $r$ from a site is defined
as a thin connected area composed of all the triangles that are attached to
the links located at $r$-steps from the marking site.
The number of triangles of this area and the number of disconnected
boundaries precisely correspond to $L_b(r)$ and $N_b(r)$ respectively.
It is also possible to define the geodesic distance on the dual lattice.
\par
We parametrize the above mentioned three quantities as follows:
\item{(A)}
$N_b(r) \ \equiv \ \VEV{ \hbox{number of boundaries at the step $r$} }
        \ \sim   \ r^\alpha$,
\item{(B)}
$L_b(r) \ \equiv \ \VEV{ \hbox{total length of boundaries at the step $r$} }
        \ \sim   \ r^\beta$,
\item{(C)}
$V_t(r) \ \equiv \ A(r) \ \equiv \ \VEV{ \hbox{number of triangles within
$r$ steps} }
        \ \sim   \ r^\gamma$.
\par
{\noindent}$\alpha$, $\beta$, and $\gamma$ may be called the fractal
dimensions of internal space-time geometry if they become constant in the
large $r$ asymptotic region.
If the surface is smooth, these quantities behave as follows:
$N_b(r) \sim 1,~L_b(r) \sim r,~V_t(r) \sim r^2$.
Thus deviations from these behaviors signal the fractal nature of quantum
gravitational fluctuation of space-time.
In other words we expect that the space-time surface is smooth classically
while the fractal nature of the space time is an essential feature of
quantum gravity.
\endpage
\vskip 1 cm
\noindent
\fourteenrm
3.~ $c=-2$ Model and Recursive Sampling Method
\rm
\par
In the numerical simulation of dynamical triangulation there are essentially
two methods to generate the typical surface of two-dimensional quantum
gravity : 1) flipping method by Monte Carlo and 2) recursive sampling method
which is first proposed by Agishtein and Migdal\NPrefmark{\AM} for pure
gravity ($c=0$ model).
The second method can generate much larger number of triangles than the first
method.
The second method, however, necessitates an analytic formula to generate a
typical surface and is restricted to particular models such as $c=0$ and
$c=-2$ models.
Here we investigate $c=-2$ model which has much simpler analytic formula than
$c=0$ model and is thus easier to simulate larger number of triangles:
number of triangles = $5 \times 10^6$ for $c=-2$ while $1.3 \times 10^5$ for
$c=0$.
\par
The $c=-2$ model or equivalently the two-dimensional gravity coupled to
$c = - 2$ matter was introduced and solved analytically in [\Kostov].
The partition function of two-dimensional gravity coupled to $c = -2$ matter
is given by
%%%%%%%%%%%%%%%%%%%%%%%%%%%%%%% Equation %%%%%%%%%%%%%%%%%%%%%%%%%%%%%
$$ \eqalign{ \sp(2.0)
 Z(N) \ &= \ \sum_G \,  \int \, D\bar\psi_i D\psi_i \,
   \exp \bigl\{ \, - \sum_{<ij>}
   (\bar\psi_i - \bar\psi_j)(\psi_i - \psi_j) \, \bigr\}  \cr
\sp(4.0)
       &= \  \sum_G \,  \det \Delta (G) \cr
\sp(4.0)
       &= \  {1\over{N+2}} \, T_{N+1} \, R_{N/2+1} , \cr
\sp(3.0)} \eqno(3) $$
%---------------------------------------------------------------------
where $\psi_i$ and $\bar\psi_i$ are fermion fields sitting at a site $i$ of
dynamically triangulated lattice G.
Here we fix the topology of the surface as sphere and the number of triangles
as $N$ and the area of the triangle is unity.
The fermion integration leads to a random lattice version of Laplacian
$\Delta(G)$.
In this particular model the determinant of the Laplacian and dual planar
$\phi^3$ diagrams are related.
$T_{N+1}$ is the number of rooted dual tree diagrams with $N+2$ external
legs while $R_{N/2+1}$ is a number of rainbow diagrams with $N/2+1$ lines.
A rainbow diagram can be constructed by connecting the $N+2$ external legs
of a tree diagram in such a way that none of the $N/2+1$ lines crosses over.
The third equality in eq.~(3) is a very useful one and can be established
by an application of the well-known Kirchhoff theorem. %\REFmark{\Kirchhoff}
The introduction of the fermion field effectively means an embedding of a
surface into $c = -2$ dimension.
\par
$T_n$ satisfies the Schwinger-Dyson equation:
$T_{n} = \sum_{k=1}^{n-1} \, T_{k} \, T_{n-k}$,
which can be solved by introducing a corresponding generating function of
$T_n$ and leads to the solution:
$T_n  =  {(2n \, - \, 2)! / n! \, (n-1)!}$.
$R_n$ satisfies the same type of relation as $T_n$, which leads to a
relation: $R_{n-1} = T_{n}$.
\par
The recursive sampling algorithm to generate a typical surface goes as follows.
First of all one of the great advantages of $c=-2$ model is the factorization
feature of $T_{N+1}$ and $R_{N/2+1}$ in Eq.(3).
In order to generate a typical surface we independently generate a tree
diagram and a rainbow diagram with a correct probability and then connect
them.
The branching probability to divide a tree diagram with $n+1$ legs into two
different tree diagrams with $k+1$ external legs and $n-k+1$ external legs is
given by $W( n , \, k ) = { T_k \, T_{n-k} / T_n }$.
A typical tree diagram with $N$ vertices can be generated by applying $N$
times of the probability formula $W( n , \, k )$.
A typical rainbow diagram can be generated similarly.
\vskip 1cm
\noindent
\fourteenrm
4.~ Numerical Results on Fractal Structures of $c=-2$ Model
\rm
\par
We show the numerical results of the fractal dimensions $\alpha$, $\beta$
and $\gamma$ with the fittings of (A) $N_b(r) \sim r^\alpha$, (B)
$L_b(r) \sim r^\beta$, and (C) $V_t(r) \equiv A(r) \sim r^\gamma$
in Figs.1, 2, and 3, respectively.
We have measured those quantities for the following number of triangles:
(1)~$8 \times 10^3$, (2)~$4 \times 10^4$, (3)~$2 \times 10^5$, (4)~$10^6$,
and (5)~$5 \times 10^6$.
The values $\alpha$, $\beta$ and $\gamma$ are approaching to constant
values 2.55 $\pm$ 0.1, 2.53 $\pm$ 0.1 and $\sim$ 3.5, respectively, and thus
show clear fractal behaviors.
We obtain an approximate relation $\gamma - \beta \simeq 1$ numerically,
which is expected from the relation $dV_t(r)/dr \simeq L_b(r)$.
What is interesting and unexpected is that the two-dimensional quantum
gravitational space time is very much branching.
For example the space-time surface with $c=-2$ matter splits into
$\sim$6000 branches after $\sim$60 geodesic steps in the case of
$5 \times 10^6$ triangles.
See Fig.1(a).
\par
It happens that the fractal dimension of the number of boundaries
($\alpha$) is approximately same as that of the total length of the
boundaries ($\beta$).
To estimate the average perimeter length of a branch we have measured
the ratio (D): $L_b(r) / N_b(r)$ in Fig.4.
It shows clear constant behavior as a function of the geodesic distance.
The numerical value is approximately $\sim$15 in the case of $10^6$ triangles.
It looks as if all the branches have similar perimeter length.
In order to see if this is the case we have measured (E): the number of
boundaries $P_r(l)$ with a given boundary length $l$ as a function of the
boundary length $l$ for a given geodesic steps $r$ measured from a marking
point.
The results are shown in Fig.5.
As we can see in the figures, different sizes of boundary length are
uniformly distributed with roughly a power law behavior of $\sim l^{-1.8}$.
In other words there is no particular preference of the perimeter length
even though the average perimeter length is $\sim 15$.
\par
We can now imagine how the two-dimensional quantum gravitational space time
looks like.
It has many branches whose perimeter length varies small size to large size.
Number of the branches with large perimeter size is small but there exists
such branches with a certain probability.
If we look into this quantum gravitational universe microscopically, we feel
like being in a dense jungle and the universe looks like complete chaos.
The universe, however, have beautiful fractal structures and thus has an
order as a whole.
We have given here a numerical results of $c=-2$ model.
In $c=0$ model Agishtein and Migdal found a similar branching behavior of the
space-time surface, but didn't observe the fractal behavior.\NPrefmark{\AM}
This suggests that the branching behavior of the quantum gravitational space
time is quite universal phenomena irrespective of the choice of matter.
\par
The geodesic distance so far has been measured on the triangulated lattice.
Geodesic distance on the dual lattice or equivalently on the dual planar
$\phi^3$ diagram is defined as a number of minimum steps located between
two vertices.
It is one of interesting questions whether the numerical results of the
fractal structure change depending on the different definitions of the
geodesic distance.
We show the numerical result of the fractal dimension of boundary length
(F): $\beta(r) \equiv d \log L_b (r)/ d \log r$ measured with the geodesic
distance on the dual lattice in Fig.6, which should be compared with the
corresponding result in Fig.2.
As we can see, the behavior to approach the stationary value is very slow in
the case of geodesic distance on the dual lattice.
This may be understood by recognizing the fact that the geodesic distance
on the dual lattice generally takes much more steps than those on the
triangulated lattice to measure the same distance.
It is, however, natural to expect that $\beta$ defined on the dual lattice
will eventually approach to the same value as that defined on the
triangulated lattice.
\vskip 1cm
\noindent
\fourteenrm
5.~Diffusion Equation on the Dynamically Triangulated Lattice and the
Continuum Limit
\rm
\par
In this section we try to investigate the fractal structure of quantum
gravity analytically by diffusion equation.
We first define adjacency matrix $K_{ij}$.
For a given typical surface $G_0$ we number the sites of the triangulated
lattice.
Then the ($i,j$) component of the adjacency matrix $K_{ij}$ is one
if $i$-th site and $j$-th site are connected by a link as next neighboring
sites, and zero if they are not connected by a single link.
It is interesting to note that ($n,n_0$) component of $(K^T)_{n,n_0}$ counts
the number of possible random walks reaching from a marking site $n_0$ to
a site $n$ after $T$ steps.
The Laplacian defined on the dynamically triangulated lattice is given by
%%%%%%%%%%%%%%%%%%%%%%%%%%%%%%% Equation %%%%%%%%%%%%%%%%%%%%%%%%%%%%%
$$ \eqalign{ \sp(2.0)
 \Delta_L \ = \ 1 \, - \, S,~~~~~ S_{ij} \ = \ {1\over q_i}K_{ij},
\cr
\sp(3.0)} \eqno(4) $$
%---------------------------------------------------------------------
where $q_i$ is called coordination number and denotes a number of links
connected to the site $i$.
$S_{ij}$ is thus a probability of one step random walk from the site $j$
to the neighboring site $i$.
The diffusion equation on a given surface $G_0$ with $N$ triangles is now
given by
%%%%%%%%%%%%%%%%%%%%%%%%%%%%%%% Equation %%%%%%%%%%%%%%%%%%%%%%%%%%%%%
$$ \eqalign{ \sp(2.0)
\partial_T \Psi_N^{(G_0)}(T;n,n_0) \ = \ \Delta_L(G_0) \,
\Psi_N^{(G_0)}(T;n,n_0),
\cr
\sp(3.0)} \eqno(5) $$
%---------------------------------------------------------------------
where $\partial_T$ is a difference operator in $T$ direction and
$\Psi_N^{(G_0)}(T;n,n_0)$ is a wave function of the diffusion equation
and denotes the probability of finding the random walker at the site $n$
after $T$ steps from the starting site $n_0$.
A solution of the diffusion equation can be easily obtained as
$\Psi_N^{(G_0)}(T;n,n_0) = e^{T\Delta_L(G_0)}(\delta_{n,n_0})$,
where $(\delta_{n,n_0})$ is $N$-component vector with unit $n_0$ entry.
\par
We now consider the continuum limit of this diffusion equation.
First of all we recover the lattice constant $a$.
In taking continuum limit, the total physical area $A=a_i^2N_i$ is fixed
and $a_i\rightarrow 0$ ($N_i\rightarrow \infty$) is taken, where $N_i$ is
the number of triangles and $a_i^2$ is the area of a triangle.
In each step of the limiting process we select a typical surface $G_i$ for
the given number of triangles $N_i$, on which the lattice Laplacian
$\Delta_L(G_i)$ of Eq.(4) is defined.
Now the lattice version of the diffusion equation (5) can be rewritten as
%%%%%%%%%%%%%%%%%%%%%%%%%%%%%%% Equation %%%%%%%%%%%%%%%%%%%%%%%%%%%%%
$$ \eqalign{ \sp(2.0)
{1 \over a_i^2} \{ \Psi_A^{(G_i)}(T+a_i^2;x,x_0) \, - \,
                   \Psi_A^{(G_i)}(T;x,x_0) \} \ = \
{1 \over a_i^2} \Delta_L(G_i) \Psi_A^{(G_i)}(T;x,x_0) \},
\cr
\sp(3.0)} \eqno(6) $$
%---------------------------------------------------------------------
where the location of the site $x$ is measured with respect to the lattice
constant $a_i$.
Thus we identify the dimension of $T$ as that of area: $dim[T] =  dim[A]$.
In the continuum limit the solution of the diffusion equation (6) is
expected to approach the continuum wave function:
$\Psi_A^{(G_i)}(T;x,x_0) \rightarrow \Psi_A^{(G_{\infty})}(T;x,x_0)$.
Numerically we approximate the limiting surface as the typical surface($G_0$)
of the maximum size: $G_\infty \simeq G_0$
As we have already noted in Eq.(2), the metric integration is effectively
carried out for the equation (6) since we have chosen a typical surface.
This means that the quantum effect is included for the wave function of Eq.(6).
On the other hand the solution of the continuum counterpart of the diffusion
equation:
$\partial_\tau \Psi(\tau;x,x_0) = \Delta(g) \Psi(\tau;x,x_0)$
is still background metric dependent in general.
Furthermore the dimensions of $T$ and $\tau$ may not necessarily be equal.
\vskip 1cm
\noindent
\fourteenrm
5.~Fractal Dimensions by Liouville Theory
\rm
\par
An analytical treatment of fractal dimensions by Liouville theory has been
first given by Kawai and Ninomiya.\NPrefmark{\KN}
In their treatment fermion is introduced as a test particle to derive the
fractal dimension.
It has been recognized that there are several ways of defining fractal
dimensions, which may explain the discrepancy between the theoretical and
numerical results.\NPrefmark{\KKSW}
Here we derive fractal dimensions by investigating the gravitational random
walks with the help of diffusion equation and Liouville theory.
In this section we briefly sketch the derivation of the fractal dimension
while the details of the derivation will be given elsewhere.\NPrefmark{\KSW}
\par
Let us now define the comeback probability of random walk on the triangulated
lattice and relate it with the continuum expression of Liouville theory as
follows:
%%%%%%%%%%%%%%%%%%%%%%%%%%%%%%% Equation %%%%%%%%%%%%%%%%%%%%%%%%%%%%%
$$ \eqalign{ \sp(2.0)
G(T) \ &\equiv \ \Psi_A^{(G_0)}(T;x_0,x_0) \ \simeq \
{<\int dx \sqrt g \, \Psi(\tau;x,x)>_A \over <\int dx \sqrt g>_A} \
\cr
\sp(8.0)
&= \ {1 \over A}<\int dx \sqrt g \, e^{\tau \Delta} \, \Psi(0;x,x)>_A
 \ \sim \ {1 \over A} \ \sim \ {1 \over T},
\cr
\sp(3.0)} \eqno(7) $$
%---------------------------------------------------------------------
where $<O>_A$ is an expectation value of $O$ with the partition function
given by Eq.(1).
We should remind of the fact that the metric integration is effectively
carried out since we have chosen the typical surface $G_0$ for the wave
function of the comeback probability.
The initial wave function can be formally written as
$\Psi(0;x,x) = \lim_{y \rightarrow x}  \delta(y-x)1/\sqrt g$
and need to be reguralized.
The first similarity relation of Eq.(7) can be understood from the
relation: $\int dy \sqrt g \, \Psi(\tau;y,x) \, = \, 1$, which is obvious
from the provability interpretation of the wave function.
The dimensional arguments coming from the equation (6) supports the last
similarity relation.
Here comes a prediction of comeback provability (G): $G(T)T \sim const.$.
\par
We next consider how to accommodate the Weyl invariance into the diffusion
equation of random walk by using DDK\NPrefmark{\KPZDDK} formulation of
Liouville theory.
Let us consider the following quantity by Liouville theory:
%%%%%%%%%%%%%%%%%%%%%%%%%%%%%%% Equation %%%%%%%%%%%%%%%%%%%%%%%%%%%%%
$$ \eqalign{ \sp(2.0)
<\int dx \sqrt g \, \Psi(\tau;x,x)>_A \ = \
&<\int dx \sqrt g \, \Psi(0;x,x)>_A \ + \
\cr
&\tau \, <\int dx \sqrt g \, \Delta \, \Psi(0;x,x)>_A \ + \ \cdots,
\cr
\sp(3.0)} \eqno(8) $$
%---------------------------------------------------------------------
where the solution of the diffusion equation is expanded by $\tau$.
\par
Taking a conformal gauge $g_{\mu\nu}(x) = \hat g_{\mu\nu}e^{\phi(x)}$ and
introducing DDK arguments, we can rewrite the first and second terms of
the Eq.(8) as
%%%%%%%%%%%%%%%%%%%%%%%%%%%%%%% Equation %%%%%%%%%%%%%%%%%%%%%%%%%%%%%
$$ \eqalign{ \sp(2.0)
<\int dx \sqrt g \, \Psi(0;x,x)>_A \ &= \
<\int dx \sqrt {\hat g} \,
\big[{1 \over \sqrt {\hat g}}\delta(x\, - \, x_0)\big]_{x=x_0}>_A,
\cr
<\int dx \sqrt g \, \Delta \, \Psi(0;x,x)>_A \ &= \
<\int dx \sqrt {\hat g} \, \big[\overrightarrow {\hat \Delta_x} \,
e^{\alpha_{-1}\phi} \,
{1 \over \sqrt {\hat g}}\delta(x\, - \, x_0)\big]_{x=x_0}>_A,
\cr
\sp(3.0)} \eqno(9) $$
%---------------------------------------------------------------------
where the term $e^{\alpha_{-1}\phi}$ is introduced to keep the Weyl
invariance of the second term.
The expectation value $<O(g)>_A$ is now rewritten by using the well-known
expression
%%%%%%%%%%%%%%%%%%%%%%%%%%%%%%% Equation %%%%%%%%%%%%%%%%%%%%%%%%%%%%%
$$ \eqalign{ \sp(2.0)
<O(g)>_A \ =  \
\int {\cal D}_{\hat g} \phi
 &Z_{\rm{FP}} [ \hat g ]  Z_{\rm M} [ \hat g ] \,
 \delta \Bigl( \int \! d x \sqrt{\hat g} \, e^{\alpha_{1} \phi}
               \, - \, A \Bigr)\, O(\hat g, \phi) \,
\cr
&\exp \bigl(  {D - 25 \over 48 \pi}
S_{\rm L} [ \phi , \hat g ] \bigr), \cr
\sp(3.0)} \eqno(10) $$
%---------------------------------------------------------------------
where $Z_{\rm{FP}}$ is the Fadeev Popov contribution and
$S_{\rm L} [ \phi , \hat g ]$ is the Liouville action.
$\alpha_n$ appeared in Eqs.(9) and (10) is given by
%%%%%%%%%%%%%%%%%%%%%%%%%%%%%%% Equation %%%%%%%%%%%%%%%%%%%%%%%%%%%%%
$$ \eqalign{ \sp(2.0)
 \alpha_{n} \ = \
 { 2 n \over 1 \, + \, \sqrt{ (25 - c - 24n) / (25 - c) } }   .  \cr
\sp(3.0)} \eqno(11) $$
%---------------------------------------------------------------------
Invariance of the expectation value under the translation of the conformal
field $\phi \rightarrow \phi - ln\lambda/\alpha_1$ leads to the
change: \nextline
$ \delta \Bigl( \int \! d x \sqrt{\hat g} \, e^{\alpha_{1} \phi}
\, - \, A \Bigr)\, \rightarrow \,
\lambda \delta \Bigl( \int \! d x \sqrt{\hat g} \,
e^{\alpha_{1} \phi} \, - \, \lambda A \Bigr)$, which is then interpreted
as the scale change of the physical area $A \rightarrow \lambda A$.
To require the invariance of each term of Eqs.(8) and (9) under this scale
transformation $\tau$ should scale as
$\tau \rightarrow \lambda^{-{\alpha_{-1} \over \alpha_1}}\tau$.
In other words,
%%%%%%%%%%%%%%%%%%%%%%%%%%%%%%% Equation %%%%%%%%%%%%%%%%%%%%%%%%%%%%%
$$ \eqalign{ \sp(2.0)
\hbox{\rm dim}{ \tau } \ = \
\hbox{\rm dim}{  A^{ - { \alpha_{-1} \over \alpha_{1} } } } . \cr
\sp(3.0)} \eqno(12)$$
%---------------------------------------------------------------------
\par
We now point out that the expectation value of the mean squared geodesic
distance is evaluated by the standard continuum treatment
%%%%%%%%%%%%%%%%%%%%%%%%%%%%%%% Equation %%%%%%%%%%%%%%%%%%%%%%%%%%%%%
$$ \eqalign{ \sp(2.0)
\int dx \sqrt g \, \big\{ r(x,x_0) \big\}^2 \, \Psi(0;x,x) \ = \
-\, 4\tau \ + \ O(\tau^2),
\cr
\sp(3.0)} \eqno(13) $$
%---------------------------------------------------------------------
which is now related with the quantum version of the mean squared geodesic
distance in the small $\tau$ region
%%%%%%%%%%%%%%%%%%%%%%%%%%%%%%% Equation %%%%%%%%%%%%%%%%%%%%%%%%%%%%%
$$ \eqalign{ \sp(2.0)
<r^2> \
&\equiv \ \sum_x \, \{r(x,x_0)\}^2 \, \Psi_A^{(G_0)}(T;x,x_0)
\cr
&\simeq {<\int dx \sqrt g \, \int dx_0 \sqrt g \, \{r(x,x_0)\}^2 \,
\Psi(\tau;x,x)>_A \over <\int dx \sqrt g>_A} \
\cr
& \sim \ \tau \ \sim \ A^{-{\alpha_{-1} \over \alpha_1}} \ \sim
\ T^{-{\alpha_{-1} \over \alpha_1}},
\cr
\sp(3.0)} \eqno(14) $$
%---------------------------------------------------------------------
where the geodesic distance on the dynamically triangulated lattice is same
as that of the numerical simulation.
The last three similarity relations are due to the dimensional arguments.
We thus obtain an analytic prediction of the $T$ dependence of the mean
squared geodesic distance (H): $<r^2> \sim T^{-{\alpha_{-1} \over \alpha_1}}
\equiv T^{2\over \gamma(c)}$.
\par
{}From the dimensional arguments of Eq.(14), we obtain the following relation:
$dim[A] \, = \, dim[r^{-2\alpha_1/\alpha_{-1}}] \,
\equiv \, dim[r^{\gamma(c)}]$.
In the numerical simulation (C): $V_t(r) \equiv A(r) \sim r^\gamma$,
the fractal dimension $\gamma$ for $c=-2$ is estimated from the measurement
of the two-dimensional volume of space time.
It is then natural to expect that the dimension of $A(r)$ and that of the
total area $A$ is same.
We then obtain the analytic evaluation of the fractal dimension
%%%%%%%%%%%%%%%%%%%%%%%%%%%%%%% Equation %%%%%%%%%%%%%%%%%%%%%%%%%%%%%
$$ \eqalign{ \sp(2.0)
 \gamma(c)
 \ = \
 - \, 2 \, { \alpha_{1} \over \alpha_{-1} }
 \ = \
 2 \times { \sqrt{ 25 - c } \, + \, \sqrt{ 49 - c }  \over
            \sqrt{ 25 - c } \, + \, \sqrt{  1 - c }         } ,  \cr
\sp(3.0)} \eqno(15) $$
%---------------------------------------------------------------------
where some of the typical values of the fractal dimension are;
$\gamma(1) = \nextline
2(1\, + \, \sqrt 2)$, $\gamma(0) = 4$, $\gamma(-2) = {1\over 2}(3\, + \,
\sqrt{17}) = 3.56\cdots$, $\gamma(-\infty) = 2$.
We show the $c$ dependence of the fractal dimension $\gamma(c)$ in Fig.7.
As we can see from Fig.7, the fractal dimension varies smoothly from the
classical value $\gamma(-\infty) = 2$ to $\gamma(1) = 2(1\, + \, \sqrt 2)$
and gets imaginary for $c>1$.
\par
We now summarize the analytic predictions:
\item{(G)}
$G(T)T \ \sim \ const.$,
\item{(H)}
$<r^2>  \ = \ T^{2\over \gamma(-2)} \ \simeq \ T^{-0.56}$,
\item{(I)}
$\gamma(-2) \ = \ {1\over 2}(3\, + \, \sqrt{17}) \ = \ 3.56\cdots$.
\par
\noindent
First of all the analytic prediction of $\gamma(-2)  =  3.56\cdots$ should
be compared with the numerical value of $\gamma \sim 3.5$.
The theoretical value (0.561) has an excellent agreement with the
experimental value.
We show the numerical evaluations of (G): $G(T)T$ and the mean squared
distance of random walks with gravitational quantum fluctuations
(H): $<r^2>$ in Fig.8 and Fig.9, respectively.
As we can see, $G(T)T$ in Fig.8 show the clear constancy and thus excellent
agreement with the prediction (G).
In Fig.9 the numerical value of the power of $T$ is approaching to the
theoretical value and slightly away in the large $T$ region where finite
size effects may be important.
If we consider the accuracy of the vertical measure we may conclude that the
agreement with the theoretical prediction is excellent again.
\endpage
\vskip 1cm
\noindent
\fourteenrm
6.~Non-universal Nature of the Fractal Structure in the Continuum Limit
\rm
\par
In our recent investigation\NPrefmark{\KKMW} we have obtained analytic
formulation to evaluate the fractal structure of two dimensional quantum
gravity without matter ($c=0$).
In this analysis we have used the results of matrix model and introduced a
combinatorial consideration and then succeeded to derive a transfer matrix.
The notion of the geodesic distance explained in this manuscript played an
important role in the formulation.
\par
We briefly summarize the main conclusion of this
investigation.\NPrefmark{\KKMW}
Let us first define a continuum function $\rho(L;D)$ which is a function
of the boundary length $L$ of a loop located at the continuum geodesic
distance $D$ measured from a marking point.
$\rho(L;D)dL$ counts the number of boundaries whose boundary lengths lie
between $L$ and $L+dL$.
It is evaluated by taking the continuum limit from the transfer matrix and
disk amplitude of dynamical triangulation.
The functional form of $\rho(L;D)$ for $c=0$ model is given by
%%%%%%%%%%%%%%%%%%%%%%%%%%%%%%% Equation %%%%%%%%%%%%%%%%%%%%%%%%%%%%%
$$ \eqalign{ \sp(2.0)
        \rho(L;D)D^2 \ = \ {3\over 7 \sqrt{\pi}}(x^{-5/2} + {1\over2} x^{-3/2}
 + {14\over3} x^{1/2})
                         e^{-x},
\cr
\sp(3.0)} \eqno(16) $$
%---------------------------------------------------------------------
where $x=L/D^2$ is a scaling parameter.
Surprising fact is that the function $\rho(L;D)D^2$ is a universal function
with respect to the scaling parameter $x$.
This quantity $\rho(L;D)D^2$ for $c=0$ model has recently been measured
numerically and gets excellent agreement with the theoretical result
(16).\NPrefmark{\TY}
Fig.5 (a) and (b) are the lattice counterparts of the multiplicity function
of $c=-2$ model; (a) $\rho(L,r=18)$ and (b) $\rho(L,r=40)$, respectively,
where continuum geodesic distance $D$ is replaced by the lattice geodesic
step $r$.
In Fig.5 only the small $x$ region is shown.
\par
In order to examine the scaling property of the fractal structure, it is
convenient to introduce the following quantities:
%%%%%%%%%%%%%%%%%%%%%%%%%%%%%%% Equation %%%%%%%%%%%%%%%%%%%%%%%%%%%%%
$$ \eqalign{ \sp(2.0)
     <L^n> \ = \ \int_0^\infty dL ~ L^n \rho(L;D).
\cr
\sp(3.0)} \eqno(17) $$
%---------------------------------------------------------------------
We can then derive the fractal scaling behaviors of $c=0$ model on the
following quantities
%%%%%%%%%%%%%%%%%%%%%%%%%%%%%%% Equation %%%%%%%%%%%%%%%%%%%%%%%%%%%%%
$$ \eqalign{ \sp(2.0)
     <L^0> \ &\equiv \ N_b(D) \ \sim \ const \times D^3 a^{-3/2}
     	+const \times D a^{-1/2}
                      + const \times D^0,
\cr
     <L^1> \ &\equiv \ L_b(D) \ \sim \ const \times D^3 a^{-1/2}
                      + const \times D^2,
\cr
     <L^n> \ &\sim \ const \times D^{2n} ~~~~~(n\geq2),
\cr
\sp(3.0)} \eqno(18) $$
%---------------------------------------------------------------------
where $a$ is the lattice constant and $N_b(D)$ and $L_b(D)$ are the same
quantities as those defined in (A): $N_b(r) \sim r^\alpha$ and (B):
$L_b(r) \sim r^\beta$ in the numerical simulations except that the argument
is now the continuum geodesic distance $D$.
\par
As we can see from Eq.(18), $N_b(D)$ and $L_b(D)$ include the inverse power
of the lattice constant dependent part as a dominant contribution.
For this $c=0$ model the fractal dimension $\gamma = \beta + 1 = 4$ obtained
from the $D$ dependence of $L_b(D)$ coincides with that of $\gamma(0) = 4$
obtained from the formula (15).
What is unexpected is that the coefficient of this power dependent term is
lattice constant dependent and thus has non-universal nature.
It should also be noted that the fractal dimension obtained from
$<L^2> \sim D^4$, which has the same dimension as $A(D) \sim D^\gamma$,
happens to reproduce the same fractal dimension $\gamma = 4$ discussed above.
This fractal dimension derived from the $<L^2>$ does not have lattice
constant dependence and thus should have a universal nature.
It is a crucial question if $\gamma(0)=4$ obtained from the Liouville theory
is equivalent either with the one derived from $<L^1>=L_b(D)$ or with another
one derived from $<L^2>$.
\par
The analytic result of $c=0$ model given in Eq.(18) suggests that $N_b(D)$
and $L_b(D)$ include non-universal part as a dominant contribution which
become irrelevant in the continuum limit but show a fractal nature in the
microscopic level.\par
Agishtein and Migdal have carried out the numerical simulation
for $c=0$ model\NPrefmark{\AM} and measured the quantity
$\gamma (r)= d\log A(r)/d\log r$ with the maximum number of triangles
$\simeq 1.3 \times 10^5$.
The observed fractal dimension at the maximum number of traingles is
roughly 3 and still far below the analytically expected value 4.
We believe that $\gamma(r)$ will approach to the analytic value for much
larger number of triangles.

\vskip 1cm
\noindent
\fourteenrm
7.~Conclusion and Discussions
\rm
\par
In this manuscript we have reported the following numerical results and
analytic predictions and corresponding figures:
\item{(A)}
$N_b(r) \ \equiv \ \VEV{ \hbox{number of boundaries at the step $r$} }
        \ \sim   \ r^\alpha$ ---------- Fig.1,
\item{(B)}
$L_b(r) \ \equiv \ \VEV{ \hbox{total length of boundaries at the step $r$} }
        \ \sim   \ r^\beta$ ---------- Fig,2,
\item{(C)}
$V_t(r) \ \equiv \ A(r) \ \equiv \ \VEV{ \hbox{number of triangles
within $r$ steps} }  \ \sim   \ r^\gamma$ ----------Fig.3,
\item{(D)}
 $L_b(r) / N_b(r)$ ---------- Fig.4,
\item{(E)}
 The number of boundaries $P_r(l)$ with a given boundary length $l$ as a
 function of the boundary length $l$ for a given geodesic step $r$ measured
 from a marking point ---------- Fig.5,
\item{(F)}
$L_b(r)  \ \sim  \ r^\beta$, where $r$ is defined on the dual
lattice ---------- Fig,6,
\item{(G)}
$G(T)T \ \sim \ const.$ ----------Fig.8,
\item{(H)}
$<r^2>  \ = \ T^{2\over \gamma(-2)} \ \simeq \ T^{-0.56}$ ----------Fig.9,
\item{(I)}
$\gamma(-2) \ = \ {1\over 2}(3\, + \, \sqrt{17}) \ = \ 3.56\cdots$ ----------
Fig.3.
\par
The numerical results (A), (B) and (C) show clear fractal structure of
two-dimensional quantum gravitational space-time.
In particular the space time is violently branching.
The results (D) and (E) show that the perimeter length of the branches
varies from small to large sizes but the average size is independent of the
geodesic distance measured from a marking point.
This suggests an existence of some rule even for $c=-2$ model based on the
fractal nature of the branching behavior .
In fact for pure gravity ($c=0$) we have analytically derived the $c=0$
counterpart of the multiplicity function (E) of Fig.5.
The result (F) in Fig.6 shows that the geodesic distance defined of the dual
lattice enforces the very slow approach of the fractal dimension to the
asymptotic value.
The analytic results (G) and (H) have excellent agreement with the numerical
results of corresponding figures Fig.8 and Fig.9, respectively.
Thus the two-dimensional quantum gravity with dynamical triangulation can
be treated by the diffusion equation of random walk with quantum gravitational
fluctuations.
The lattice version of the wave function and the continuum counterpart of
diffusion equation are related by Eq.(7), which makes it possible to
accommodate the Liouville theory.
The fractal dimension (I): $\gamma(-2)$ obtained from the formula (15)
excellently agrees with the numerical results of $c=-2$ model obtained
from Fig.2 and Fig.3 while the $\gamma(0)$ obtained from (15) coincides
with the analytic results derived from (18).
Therefore we tend to believe that the formula (15) derived from the
Liouville theory provides correct fractal dimensions of two-dimensional
quantum gravity with matter fields.
\par
As we have mentioned in the last section, $N_b(D)$ and $L_b(D)$ of $c=0$
model include non-universal lattice constant dependent part as a dominant
power contribution of geodesic distance which becomes irrelevant in the
continuum limit.
This may also be the case for $c=-2$ model.
If $N_b(D)$ and $L_b(D)$ include non-universal part even for $c=-2$ model,
we are puzzled why the fractal dimension $\gamma \simeq \beta + 1 \simeq 3.5$
obtained from the numerical results of $L_b(r)$ and $A(r)$ excellently agrees
with the analytic result $\gamma(-2) \ = \ 3.56\cdots$ obtained from the
formula (15) which correctly reproduces the $c=0$ analytic result.
This kind of question may be partially answered by numerical measurements
for the models other than $c=0$.
Recently we have measured $\rho(L;D)$ and $<L^n>$ for $c=-2$ model
numerically and observed the similar scaling behavior as $c=0$
model.\NPrefmark{\KKMSW}
The problems are not yet completely settled at this moment since we have
not yet obtained the analytic results of $<L^n>$ for the model other
than $c=0$.
There are still several other issues to be cleared up: in particular the
relation between the derivations of fractal dimensions by Liouville theory
and the exact treatment.

%
%
%%%%%%%%%%%%%%%%%%%%%%%%%%% Acknowledgments %%%%%%%%%%%%%%%%%%%%%%%%%%
%
\vskip 1cm    % preprint version by original TEX
%\vfill\eject % submitting version
%
\ack
The author would like to thank F. David, H. Kawai, V. A. Kazakov,
T. Mogami, Y. Saeki and Y. Watabiki for useful discussions and fruitful
collaborations.
%
%The authors would like to thank Prof. ??? for many useful discussions.
%They also thank ??? for valuable discussions.
%
%%%%%%%%%%%%%%%%%%%%%%%%%%%%% Ending of program %%%%%%%%%%%%%%%%%%%%%%%%
%
\endpage
%
%%%%%%%%%%%%%%%%%%%%%%%%%%%%%%% Appendix %%%%%%%%%%%%%%%%%%%%%%%%%%%%%%%
%
%\topskip 30pt
%\appendix
%\Appendix{A}
%\vskip 10pt
%
%\noindent{ NOTATIONS }
%\par
%---------------------------------------------------------------------
%
%\vfill\eject
%
%======================================================================%
%
%%%%%%%%%%%%%%%%%%%%%%%%%%%%% Ending of program %%%%%%%%%%%%%%%%%%%%%%%%
%
%\endpage
%
%%%%%%%%%%%%%%%%%%%%%%%%%%%%% Figure Captions %%%%%%%%%%%%%%%%%%%%%%%%%%
%
\def\Figures{
\centerline{ \BIGr FIGURE CAPTIONS}
\vskip 8pt
\item{\rm Fig.~1a}{(A): $ N_b (r)$ --- $r$ dependence,
                   where $N_b (r)$ is the number of boundaries at
                   the step $r$ for various number
                   of triangles of sphere topology:
                   (1)~$8 \times 10^3$, (2)~$4 \times 10^4$,
		   (3)~$2 \times 10^5$, (4)~$10^6$, and (5)~$5 \times 10^6$. }
\item{\rm Fig.~1b} {(A): $\alpha (r) \equiv d \log N_b (r)/ d\log r $
                   for various number of triangles,
	           where $\alpha (r)$ is the fractal dimension parametrized
                   in (A). }
\item{\rm Fig.~2}{(B): $\beta (r) \ \equiv d \log L_b (r)/ d \log r$
                   for various sizes of triangles, where the fractal
                   dimension $\beta(r)$ is parametrized in (B). }
\item{\rm Fig.~3}{(C):$\gamma (r) \ \equiv d \log V_t (r)/ d\log r$
                   for various sizes of triangles, where $\gamma(r)$
                   is parametrized in (C). }
\item{\rm Fig.~4}{(D): $L_b(r)/N_b(r)$ for the case of
                   $10^6$ triangles.}
\item{\rm Fig.~5}{(E): The number of boundaries $P(l)$ with a given boundary
                   length $l$ as a function of the boundary length $l$
                   for the geodesic distance (a) $r=18$ and (b) $r=e=40$,
                   where the number of triangles is $10^6$.}
\item{\rm Fig.~6}{(F): $\beta (r) \ \equiv d \log L_b (r)/ d \log r$
                   for various sizes of triangles,
                   where the geodesic distance here is defined on
                   the dual lattice.}
\item{\rm Fig.~7}{ The fractal dimension $\gamma (c)$ given by Eq.(15)
                   as a function of the matter central charge $c$.}
\item{\rm Fig.~8}{(G): $G(T)T$ as a function of geodesic step $T$,
                   where $G(T)$ is the comeback probability of random walk
		   given by Eq.(7) for the case of $10^6$ triangles.}
\item{\rm Fig.~9}{(H): $\delta \equiv d \log <r^2>/ d\log T$ as a function
                        of geodesic step $T$, where $<r^2>$ is the mean
                   squared geodesic distance. The theoretical value $0.561$ is
                   shown as a solid line in the figure. }

}
%
%%%%%%%%%%%%%%%%%%%%%%%%%%%% Table Captions %%%%%%%%%%%%%%%%%%%%%%%%%%%
%
%\def\Table{
%
%\centerline{\BIGr TABLE CAPTION}
%
%\vskip 8pt
%
%\item{\rm Table~1 }{}
%
%          }
%
%======================================================================%
%
\par \penalty-400 \vskip\chapterskip
%  \spacecheck\referenceminspace \closeout\referencewrite
% 9/24/1986 The above 1 line was changed as below by H.Mawatari
   \spacecheck\referenceminspace \immediate\closeout\referencewrite
   \referenceopenfalse
   \line{\fourteenrm\hfil REFERENCES\hfil}\vskip\headskip
   \input reference.aux

\vfill\eject
\Figures
%
%\vskip 48pt
%
%\Table
%
\vfill
%======================================================================%
%
\bye